\documentclass{IOS-Book-Article}

\usepackage{mathptmx}
\usepackage{soul}\setuldepth{article}
\usepackage{booktabs}
\usepackage{textcomp, gensymb}
\usepackage{footmisc}
\usepackage{graphicx}
\usepackage[numbers]{natbib}

\usepackage{hyperref}

\def\hb{\hbox to 11.5 cm{}}

\begin{document}

\pagestyle{headings}
\def\thepage{}
\begin{frontmatter}              

\title{Seeking Information with RAG-Assistants: Does Model Size Matter in Human-AI Collaborations?
}
\author[A]{\fnms{Lennard C.} \snm{Froma}\orcid{0009-0006-2061-9208}%
\thanks{Corresponding Author: Lennard Froma, l.c.froma@liacs.leidenuniv.nl}},
\author[A]{\fnms{Tom} \snm{Kouwenhoven}\orcid{0000-0003-2480-4073}},
\author[B]{\fnms{Maaike H.T.} \snm{de Boer}\orcid{0000-0002-2775-8351}},
\author[A,C]{\fnms{Catholijn M.} \snm{Jonker}\orcid{0000-0003-4780-7461}}, and
\author[A]{\fnms{Max J.} \snm{van Duijn}\orcid{0000-0003-0798-9598}}

\runningauthor{L.C. Froma et al.}
\address[A]{Leiden University - The Netherlands}
\address[B]{TNO - The Netherlands}
\address[C]{TU Delft - The Netherlands}

\markboth{}{\today\hb}



\begin{abstract} 
Much research on LLMs has focused on increasing benchmark performance. However, the evaluation of such models in real-world collaborative human–AI workflows has stayed behind. This work evaluates a chatbot-style assistant based on Retrieval-Augmented Generation (RAG) in a realistic multi-turn information-seeking scenario inspired by workplace settings where compliance with local legislation and secure handling of sensitive data are often key. Specifically, we examine the performance of humans (N=112) assisted by RAG-assistants compared to LLM-only or LLM+RAG baselines. In this setting, we investigate how underlying model size (3B, 8B, and 70B) shapes the human-AI collaborative dynamic and how it influences perceived usability and satisfaction. 
Results show that the performance gain of human-AI collaboration over the model-only baselines is significant, irrespective of model size, suggesting that hybrid systems are beneficial in information-seeking scenarios. Interestingly, however, perceived usability and satisfaction among participants showed little difference across model sizes.
This demonstrates a nuanced trade-off between model size, performance, and user perception. Our work highlights the added value of evaluating AI applications in actual multi-turn interactions with human users, looking at usability and satisfaction besides accuracy, rather than focusing on benchmark performance only.
\end{abstract}

\begin{keyword}
Human-AI Interaction, User Experiment, Multi-Turn Interaction, Large Language Model, Retrieval Augmented Generation
\end{keyword}
\end{frontmatter}
\markboth{\today\hb}{\today\hb}
\section{Introduction}
Large Language Models (LLMs) are increasingly used in professional areas where information processing is essential \cite{chang2024survey}. 
In this paper, we focus on context-aware question answering (QA) and information seeking in realistic workplace scenarios.
Professionals are often tasked with analysing large volumes of documentation.
While LLMs hold potential to offer support in this process, their deployment is constrained in cases where data is sensitive and compliance with, for example, the EU AI Act is required. 
To address such constraints, open, transparent, privacy-respecting, and locally deployable language models are needed \cite{barbereau2024gpt}. 
This poses the question of how small such models can be in human-AI collaboration, while maintaining adequate performance and usability. 
 
Evaluation of LLM applications, including in QA, often focuses on established benchmarks, relying mainly on quantitative metrics such as accuracy, precision, recall, or similarity scores to assess performance \cite{asai2024selfrag, NEURIPS2024_db93ccb6,wang2025speculative}. 
For many of these problems, benchmark evaluation is well-suited as responses typically fall into discrete categories \cite{arslan2024survey}.
However, benchmark approaches often favour larger commercial models instead of more versatile, locally deployable ones \cite{10.5555/3737916.3738251, lyu2025crud}.  

Qualitative assessments, while less common, often highlight dimensions such as fluency, reliability, and perceived single-turn quality, where interactions consist of one question as input, followed by the assistant's response \cite{li2025matching}. 
However, professional users typically engage in collaborative task solving spanning multiple turns between model output and human input \cite{akata2020research}. 
This means that interpretive nuance and interactive dynamics over multiple turns become significantly more important, and that usability and user perception become key factors influencing the outcome.
The disconnect between benchmark evaluations and real-world LLM applications offers an incomplete picture.

\begin{figure*}
    \centering
    \includegraphics[width=\linewidth]{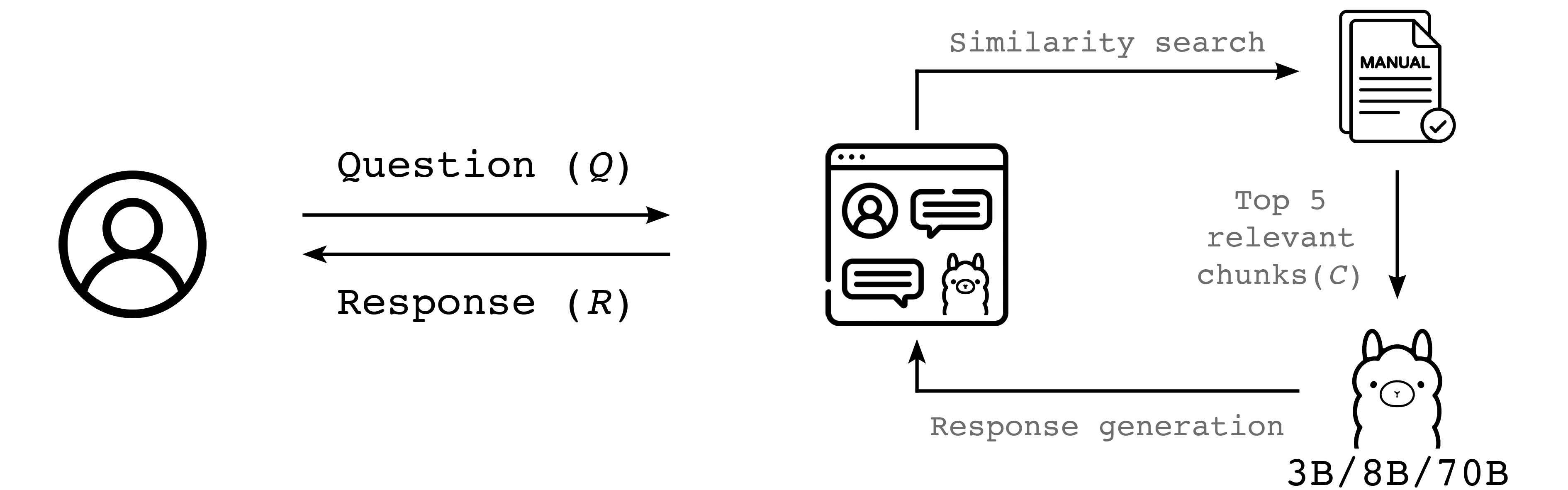}
    \caption{An overview of the RAG pipeline.}
    \label{fig:methodOverview}
\end{figure*}

Here, we address this disconnect by combining quantitative and qualitative assessments in an actual hybrid setting.
We set up an experiment in which human participants (N=112) had to answer a set of detailed questions about a large document (400+ pages), resembling a workplace information-seeking task.
Participants had direct access to a PDF of the document and a chatbot-style assistant that could be questioned about the contents. 
This assistant, henceforth called the RAG-assistant, used Retrieval Augmented Generation (RAG) in combination with open-weight generator LLMs of different sizes, as shown in \autoref{fig:methodOverview}. 
This technique resembles a realistic workplace scenario where information is kept up-to-date through external databases.
Participants could choose to answer questions either by retrieving information directly from the document, via the RAG-assistant, or via a combination of both methods. 
Financial rewards (a bonus on top of a fixed remuneration) were given for correct answers to the questions, motivating participants to complete the task accurately. 
This setup allows us to examine three main questions. 
Firstly, we gauge the added value of human-AI interaction, and ask whether the human-AI setting outperforms the LLM-only and LLM+RAG (baseline) settings (\textbf{RQ1}).
Secondly, within the human-AI setting, we ask whether model size yields higher task accuracy (\textbf{RQ2}) and whether the perceived usability and satisfaction increase with model size \textbf{RQ3}. 
Our contributions include:
\begin{enumerate}
    \item a systematic evaluation/comparison of RAG-assistants based on small, open-weight LLMs of three different sizes;
    \item a realistic scenario simulating real-world contexts;
    \item an experiment in which participants complete an information-seeking task over multiple turns, measuring accuracy, usability, and satisfaction.
\end{enumerate}

\section{Related work} 
\subsection{RAG}
LLMs are limited to the knowledge encoded during training.
This means they cannot recall information absent from their training data or fine-tuning procedures, which poses a significant obstacle to the application of LLMs when domain-specific knowledge is required \cite{kandpal2023large}.
To address this constraint, models have been equipped with the ability to consult external sources during inference, as demonstrated by WebBrain \cite{qian2023webbrain} and Web-GPT \cite{nakano2021webgpt}. 
This functionality is now present in many state-of-the-art commercial systems. 
However, working with such systems can be undesirable for a variety of reasons, including when working with sensitive data that has to be processed within local environments.
In such cases, RAG \cite{lewis2020retrieval} can be used to augment the knowledge encoded in the LLM with external, offline knowledge bases. 

At their core, RAG-assistants retrieve relevant document excerpts during inference and pass them to an LLM, which uses the provided information to formulate an answer, producing more up-to-date and/or task-specific answers \cite{lewis2020retrieval, ram-etal-2023-context}.
We adopt a sequential retrieval-augmented generation approach \cite{li2025matching}, in which relevant documents are first retrieved and then used by the generator (an LLM) to produce a response, which we found to be well-suited to our use case.

An approach to leveraging LLM capabilities in RAG is to have the model reformulate the original query to better align with the task's requirements \cite{ma2023query, dhole2024genqrensemble}.
Instead of rewriting the query, LLMs can also be used to provide feedback on the returned documents in terms of match with the posed question, and retrieve new documents if there is no match, making it self-reflective \cite{asai2024selfrag}.
A related concept is interactive information retrieval \cite{ruthven2008interactive}, in which rewriting is part of a broader set of techniques known as query reformulation \cite{10.1145/3759453}. 
Query reformulation is an important part of conversational/interactive RAG as it condenses the evolving and complex dialogue context into a more precise query.
In this work, we do not apply this technique by adding another LLM inference call, but rather let humans reformulate the queries.

Recent work has explored the viability of integrating RAG pipelines with smaller LLMs \cite{barbereau2024gpt}.
Smaller LLMs combined with RAG can meet many AI application needs more efficiently and economically \cite{belcak2025small}. 
MiniRAG \cite{fan2025minirag} introduces a lightweight, graph-based retrieval system tailored for smaller LLMs that compensates for the limited reasoning capacity of small generators while reducing the overall memory footprint by up to 75\%. 
Similarly, \citep{huang2024survey, shao2024scaling} show that smaller models often benefit disproportionately from high-quality retrieval, partially closing the performance gap to larger LLMs at the application level. 
Furthermore, \citep{shi2025models} found that larger models can sometimes be less favourable to work with in RAG applications.
Similarly, under some conditions, smaller models show better general alignment than their larger counterparts \citep{de2024evaluating}.

Besides these advantages, smaller language models in RAG applications face notable challenges.
Their smaller parameter capacity limits their ability to integrate complex or conflicting evidence retrieved from context, which can lead to hallucinations or superficial responses \cite{huang2025survey}.
Moreover, they often struggle in zero-shot generalisation and require more carefully curated prompts to perform competitively \cite{xu2022go}. 
Even under favourable conditions, models remain dependent on the placement of relevant information within the provided context \cite{liu-etal-2024-lost}, especially for smaller models. 
However, \citep{lepagnol-etal-2024-small} shows that under the right circumstances (in terms of, e.g., architecture, retriever size, data structure), smaller models can outperform bigger ones. 
This shows that smaller language models augmented with RAG hold promise for applications where local deployment is desirable.

\paragraph{Evaluating RAG} \label{subsec:evalrag}
RAG applications are generally evaluated using benchmarks that assign quantitative scores, such as ROUGE \cite{lin2004rouge}, BLEU \cite{papineni2002bleu}, accuracy, precision, recall, and error rates to compare performance. 
Examples of evaluation benchmarks are CRAG, RAGBench, TrustLLM, and RAGAs.
CRAG \cite{10.5555/3737916.3738251} is a large-scale, multi-domain factual QA benchmark designed to evaluate and stress-test RAG applications across diverse question types.
RAGBench \cite{friel2024ragbench} is a benchmark designed to evaluate RAG applications across five industry-specific domains and to address the lack of standardised, actionable evaluation methods for real-world RAG applications.
RAGAs \cite{es-etal-2024-ragas}, an evaluation framework for RAG pipelines that measures retrieval quality, faithfulness, and answer relevance without human annotations, enabling faster and more scalable assessment of RAG applications.
TrustLLM \cite{pmlr-v235-huang24x} is a benchmark and analysis framework for evaluating the trustworthiness of LLMs across multiple dimensions, including truthfulness, safety, and fairness, revealing notable trade-offs between openness, reliability, and utility. 

Evaluation, including a human perspective in RAG evaluation, has been studied.
\citep{meyer2024comparison} involved human evaluation through a 5-point scale (0, 0.25, 0.5, 0.75, 1), rating the quality of the answer. 
Although this includes human feedback, we believe it does not evaluate the complex dynamics that human-AI collaboration entails.
Others introduced an evaluation framework for assessing RAG outputs from a human perspective \cite{mangold2025human}. However, the framework does not involve direct human-AI interaction; instead, interactions are observed and evaluated.
We add a dimension to the evaluation by assessing not only objective performance but also collecting extensive feedback from participants.

\subsection{Human-AI collaboration}
Experiments have shown the added value of human-AI collaboration as it can outperform humans working in isolation \cite{noy2023experimental}.
A meta-analysis, however, showed that this is context-dependent \cite{vaccaro2024humanai}.
For decision-making tasks, human-AI collaboration tends to decrease performance, whereas for more open-ended tasks, human-AI collaboration increases performance.
Earlier work has studied collaborative LLM applications across fields, from which a taxonomy of human-LLM interaction was created \cite{10.1145/3613905.3650786}.
Examples of such systems are recommender systems \cite{schemmer2022should} and strategic planning \cite{li2025text}. 
Others focus on specific domains, such as medical \cite{tang2023medical}, legal \cite{choi2024ai,vereschak2021evaluate}, mathematics \cite{xu-etal-2024-chatglm} or coding \cite{zhang-etal-2024-codeagent}. 

In such systems, trust and perceived usefulness are key indicators for collaborative success \cite{chae2025factors,afroogh2024trust}.
The extent to which people trust LLMs is strongly influenced by perceived expertise \cite{ramrath2024trust}. 
LLM usage is also heavily task and person-dependent, as variations can lead to different strategies. \cite{zhou2024developing}.
In our setup, as part of Contribution 2 formulated above, we hypothesize that (part of the) shortcomings of using smaller language models will be mitigated through interaction with humans, in line with the general advantages of hybrid approaches \cite{akata2020research}.
\paragraph{Evaluating Human-AI collaboration}
Measuring user satisfaction combines elements like trust, perceived usefulness, and expertise as they correlate heavily \cite{davis1989technology}, and is a research field with a long tradition \citep{xie2024does}. 
Recent work focuses on satisfaction estimation in conversations.
SPUR \cite{lin2024interpretable} is successful in using LLMs to estimate user satisfaction in conversational systems. 
This estimation method assigns binary satisfaction values to prompts based on patterns learned from labelled data, crucially missing the point that user satisfaction is often nuanced and context-dependent.
Similar approaches estimate satisfaction using sentiment analysis in conversations between human and AI applications \cite{song2023speaker}. 
\citep{wang-etal-2022-understanding} took a broader empirical approach by putting out a questionnaire with general user satisfaction for all-day LLM use. 

User satisfaction in a hybrid RAG context, however, is less well studied.
Prior work in conversational RAG (LLM-LLM conversations) explored models that are too resource-intensive for local deployment \cite{katsis2025mtrag}; besides that, human judgment is only used for annotation, not for interaction.
Our work addresses this gap by specifically targeting QA/hybrid task-solving and user satisfaction for differently sized, locally deployable models.

\section{Methodology}
This study investigates the effect of model size on task performance, perceived usability, and perceived satisfaction.
We compare three conditions across two settings. 
The conditions refer to the model sizes of three instruction-tuned, smaller-sized LLMs: 3B (Llama3.2), 8B (Llama3.1), and 70B (Llama3.3) from the Llama3 family \cite{grattafiori2024llama}.
The settings are LLM+RAG and human-AI.
In the LLM+RAG setting, the original question as posed to the participant is used as a single-shot input for the RAG-assistant.
We consider this setting to be the baseline, consistent with how benchmark performance is established across the literature, as discussed in Section \ref{subsec:evalrag}.
In the human-AI setting, participants could interact with the RAG-assistant to come up with an answer.
LLM-only (i.e., querying LLMs from all conditions without RAG functionality) is also evaluated to demonstrate the suitability of the tasks presented to participants.

The generator LLM parameter size serves as the independent variable, while all other technical parameters are held constant. 
The range of sizes we chose represents practical trade-offs encountered in real-world deployments: while 70B models approximate the performance of state-of-the-art commercial models \cite{grattafiori2024llama}, their computational requirements are substantial; in contrast, 3B and 8B models are so much less demanding that they can be deployed in edge computing supporting full control over privacy-sensitive data handling. 
Additionally, 3B and 8B models contribute substantially to policies set to reduce the environmental footprint of AI usage in workplace routines.
The parameter count ranges from small (3B) to medium (8B), to large (70B), allowing us to systematically explore how model size impacts both objective performance and subjective human evaluation. 
Since our baseline method matches the process of widely used benchmark evaluations (i.e., single-turn interactions), we can gauge the added value of human-AI interaction over multiple turns by comparing the baseline with our participants' performance.

\subsection{Task}
\label{sec:task}
Participants have to answer specific questions about a publicly available flight manual of over 400 pages \footnote{For our full experimental setup, questions, and code, visit https://tinyurl.com/peerRespository\label{footnote:1}}. 
The manual serves as a realistic example of domain-specific materials due to its length, structural complexity, and mix of technical and descriptive content, mirroring the language, information density, and domain focus of texts used in many different professional areas. 

Based on the manual, nine questions were created. 
These questions varied in difficulty, determined by the amount of information required and how dispersed the information is. 
The questions varied from procedural \textit{`What should a pilot do when a door opens mid-flight?'} to practical \textit{`In the document, there are many different conventional grades of gasoline for aeroplanes mentioned. Can you tell how many there are?'}. 
All nine questions and the data used can be found in the full experimental setup \footref{footnote:1}.

We ensured that participants actually retrieved information from the document to answer the questions in three ways. 
Through pre-selection, we excluded workers with domain experience, making it unlikely that they knew the answers to the questions off the top of their heads. 
We excluded participants post-hoc who indicated that they had used external sources or tools to answer the questions (see \autoref{sec:procedure}).  
We also ruled out the possibility that the Llama models, which lack RAG functionality, could answer questions correctly through data memorisation alone.
This was done by prompting each model three times using the original questions without any further instructions or information.
The models generally failed to produce correct answers based on their internal knowledge base only. 
This indicated that the questions are suitable for our experiment and once again exposed the constraints of using such models without an additional domain-specific knowledge base. 
These results are added as LLM-only in \autoref{sec:results}.

Answers to all questions could be found in the manual, meaning there were no unanswerable or `trick' questions. 
To arrive at the right answer, participants were free to choose whether to use the source document, the RAG-assistant, or both. 
Participants could use any preferred wording of their question to prompt the RAG-assistant.

\subsection{RAG-assistant}
\label{sec:technicalDetails}
Communication between the human and the RAG-assistant follows the architecture shown in \autoref{fig:methodOverview}.
First, the source document of 406 pages is chunked ($chunksize=1024$) into 2497 chunks by the embedder $intfloat/multilingual-e5-large-instruct$ \cite{wang2024multilingual} because of its strong retrieval performance on the Massive Text Embedding Benchmark \cite{muennighoff-etal-2023-mteb}. 
Upon receiving a user question ($Q$), it is embedded. 
The cosine similarity between the embedded question and embedded chunks is calculated, after which the top 5 (similar to \citep{fan2025minirag}, \citep{turing2025small}) most relevant chunks $C$ are selected.
The top 5 is determined by the highest similarity score between the question $Q$ and the text chunks $C$.
These retrieved chunks are then passed to the generator model together with a RAG system prompt \footref{footnote:1}, giving the LLM context and instructions.

The retriever selection was based on a pre-study.
For each model size, we tested three different \textit{intfloat-e5} retrievers on their LLM+RAG performance and compared the average over three runs.
The results of evaluating multiple retrievers across all three generator models showed that the \textit{small} retriever model obtained the best overall performance with only marginal improvements over the other retrievers.
Although the \textit{small} embedding model achieved slightly stronger performance on our questions, we opted for a larger embedding model \textit{intfloat-e5-large} due to its greater representational capacity.

The generator model is assigned to the participant out of the three possible models. 
The response ($R$ in \autoref{fig:methodOverview}) is based on the information provided earlier in the conversation, the user's initial question, and the provided documents.
This response is then returned to the user.

All generator models were trained on data with the same cut-off (December 2023).
It should be noted that larger models inherently require more training data, meaning they most likely utilise more of the dataset than smaller models. 
Combined, this isolates the effect of model scale while minimising confounding factors related to model origin or alignment technique.
The model was always configured with a temperature of 1. 
This setting controls the randomness of the model’s responses, with 1 providing a balanced level of variation and coherence.

\subsection{Experimental Procedure}
\label{sec:procedure}
For each condition, we recruited 40 participants via Prolific (\textit{www.prolific.com}), yielding $N=120$ ($M_{age} = 39.4$, $\textit{SD}_{age} = 10.3$, females = $60$, and males = $60$).
From this group of participants, $8$ were removed after the experiment as they indicated external tools were used to answer the questions, leaving a total of $112$ participants who met the inclusion criteria. 
The participants were distributed across the three conditions as follows: 37 in condition 1 (3B), 37 in condition 2 (8B), and 38 in condition 3 (70B).
Ethics approval was obtained from the university's HREC (2025-011).
The average reimbursement (due to varying bonus payments) was \pounds 12.88, which is both above the prolific minimum (\pounds6-\pounds8) and the ethically recommended wage (\pounds9-\pounds12) according to Prolific \footnote{https://researcher-help.prolific.com/en/articles/445230-prolific-s-payment-principles}.

The experiment consisted of randomised controlled trials followed by cross-sectional surveys.
After obtaining informed consent, participants received instructions for the experiment and were kindly asked not to engage in other activities during the experiment.
Then participants answered two experience-related questions: one on experience with LLMs and one on experience with flying aeroplanes (real life or simulators), given the nature of the data set.

The participants were directed to a Streamlit \footnote{streamlit.io } hosted website.
This website provided access to the flight manual via a conventional chat interface, i.e., the RAG-assistant (left side) and a PDF viewer (right side).
The chat interface also showed the relevant passages from the source document used by the assistant for answer generation.
The questions were presented in the same order for all participants and could be answered using the PDF directly, the RAG-assistant, or a combination of both.
Participants were randomly assigned one of three model conditions distributed equally and randomly, making this a between-subject study. 
Participants were not informed which model their RAG-assistant was using.
After answering the question, they were asked how they arrived at this answer: through the RAG-assistant, the source data, a combination of both, or neither. 
After the experiment, participants filled out a questionnaire, which we detail below.


\subsection{Evaluation}
\label{sec:evaluation}
The answers to open questions are categorised as correct, partially correct (only applicable to multi-part questions), or incorrect. 
This was done by two independent raters, one being the main author of this paper, and the other a colleague independent of this research.
The multiple-choice questions had only one correct answer, resulting in either correct or incorrect classification.
Accuracy was calculated by assigning 1 point for correct, 0.5 for partially correct, and 0 for incorrect answers, and dividing the total score by the total possible points.
A one-sample t-test is used to compare the baseline performance (result from the base strategy) to the human-AI performance.

To analyse participants’ performance across experimental conditions, we fitted a linear mixed-effects model \cite{laird1982random}. 
Each participant responded to multiple questions, and each question was seen by multiple participants, introducing non-independence in the data. 
To address this, we included random effects for both participant (hashed Prolific ID for anonymisation) and question ID, controlling for individual differences and variability in question difficulty. 
Model size was included as a fixed effect to estimate its influence on the outcome variable: the scoring of each response. 
This approach allowed us to estimate the effect of experimental conditions on performance while appropriately compensating for possible random effects that influence performance. 
Combined, this led to the following model: $score \sim model\_size + (1 | participant_{ID}) + (1 | question_{ID})$, with Tukey's p-value adjustment for comparing a family of three (all conditions) estimates.

Our evaluation of usability and satisfaction is inspired by the ASA framework as described in \citep{fitrianie2022artificial}.
While this framework originally comprises 19 questions, we restricted our analysis to the 10 questions that best fit our specific use case and align with the scope and objectives of this study. 
The questions were rated on a 5-point Likert scale used to calculate ASA scores as described in the framework.
These ASA scores allow us to holistically compare the evaluation of the differently sized RAG-assistants.

To compare participants' responses on the questionnaire across the three experimental conditions at the per-question level, we conducted a one-way analysis of variance for ordinal data \cite{kruskal1952use}. 
This test assessed whether there were statistically significant differences in average scores between conditions. 
When the Kruskal-Wallis test indicated a significant effect, we performed post hoc pairwise comparisons using Dunn’s post hoc analysis \cite{dunn1964multiple} with Bonferroni correction \cite{abdi2007bonferroni} to identify which specific conditions differed from each other while controlling for multiple comparisons.
The exact phrasing of questions to the RAG-assistant was logged \footref{footnote:1}.
The logs were inspected in a qualitative evaluation by the experimenter.
Additionally, the average completion time is calculated and used to compare performance across different groups, providing a quantitative measure of efficiency.

\section{Results}
\label{sec:quantRes}
\label{sec:results}
\begin{figure}[t]
    \centering
    \includegraphics[width=\linewidth]{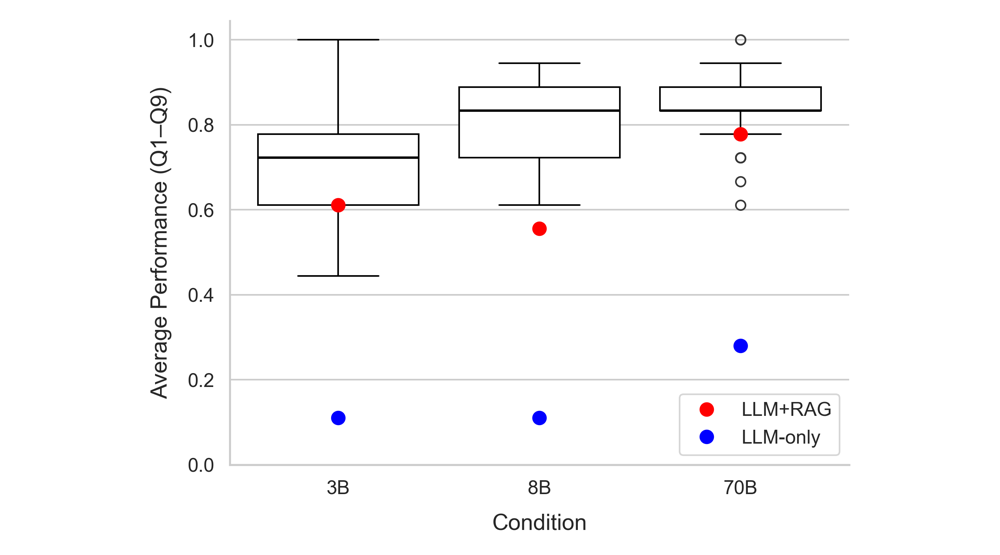}
    \caption{Performance as the accuracy of answers to the 9 test questions across conditions, compared to base-strategy (LLM+RAG) and knowledge base only setting (LLM-only). 
    }
    \label{fig:accVSBaseline}
\end{figure}

Comparing the question's evaluation from both independent raters yields a Cohen's Kappa of $.898$. 
This high Cohen's Kappa score shows little difference between raters' judgments, which is unsurprising given the unambiguous answers to the questions.

\autoref{fig:accVSBaseline} shows average accuracy scores over all questions per model. 
The box plots show the performance of the human-AI collaboration.
The blue and red dots indicate LLM-only and the LLM+RAG base strategy setting, respectively.
Adding RAG boosts performance, as indicated by the red dots, confirming the added value of including domain-specific knowledge bases.
When comparing the base setting (LLM+RAG) to the human-AI setting through a t-test, the difference in performance is significantly higher for all models ($p < .001$).
Returning to \textbf{RQ1}, this means that for all models, the human-AI setting outperformed the LLM+RAG setting.

Turning to \textbf{RQ2}, we assess performance differences across models using a mixed linear model with random effects for participants and questions.
The model showed convergence and a statistically significant increase between the 3B and 8B models ($p<.001$). 
Differences between the 70B and 8B models were not significant ($p=.354$), but performance increased significantly between the 70B and 3B models ($p<.001$).
For question answering, participants favoured the AI assistant (66\%), followed by a combination of the AI assistant and the source document (25\%), the source document alone (7\%), and none (2\%).
This shows that 91\% of the questions were generated in collaboration with the AI assistant.
No meaningful differences were found at the per-question level among groups.

Across all conditions, participants spent 25 ($\pm$13) minutes on the task.
An ANOVA test showed no significant ($p=.075$) differences between the time spent on completing the experiment.
The amount of experience with LLMs did not influence performance across conditions: $p=.612, p=.592, p=.344$ for $3B, 8B, 70B$, respectively.
Experience with flying aeroplanes (real life or simulation) showed a positive influence on performance in the 3B condition with $p=.031$; Dunn's post hoc test did not confirm this.


\begin{table*}[t]
\centering
\setlength{\tabcolsep}{6pt}
\caption{Participant ratings (mean ± standard deviation) for each question from the questionnaire across three AI models: 3B, 8B, and 70B. Bold values are the best scores. ASA score is calculated as per \citep{fitrianie2022artificial}, where higher means better, except for question 8.}
\begin{tabular}{p{7.5cm}ccc}
\toprule
\textbf{Question} & \textbf{3B} & \textbf{8B } & \textbf{70B} \\
\midrule
1. The AI assistant was good enough to assist me   & 4.4±0.6     & 4.1±1.1     & \textbf{4.6±0.6} \\
2. The AI assistant has a human-like manner        & \textbf{2.9±1.2}    & 2.6±0.9     & 2.7±1.4 \\
3. It was easy to interact with the AI assistant   & 4.6±0.8     & 4.2±1.1     & \textbf{4.7±0.6} \\
4. I can rely on the AI assistant                  & 3.8±1.0     & 3.4±1.1     & \textbf{4.2±0.9} \\
5. I see the interaction with the AI assistant as something positive & 4.3±0.7 & 3.9±0.9 & \textbf{4.6±0.6} \\
6. I like the AI assistant                         & \textbf{4.2±0.9}    & 3.8±1.0     & \textbf{4.2±0.9} \\
7. The interaction with the AI assistant captured my attention     & \textbf{4.2±1.0}    & 3.8±1.3 & 4.1±1.0 \\
8. The AI assistant has no clue of what it is doing & \textbf{1.6±0.9}    & 2.2±1.1     & 1.8±1.1 \\
9. The AI assistant is a social entity                    & 2.5±1.2     & \textbf{2.8±1.1}     & 2.4±1.1 \\
10. The AI assistant completed tasks without noticeable delays  & \textbf{4.4±0.8} & 4.2±1.0 & 4.3±0.8 \\ \midrule
ASA & 7.0 ± 5.0&4.4 ± 5.7& 7.7 ± 5.1\\
\bottomrule
\end{tabular} \hspace{2mm}
\label{tab:ai_questionnaire}
\end{table*}

\paragraph{Perceived performance} 
The subjective results of the experiment can be found in \autoref{tab:ai_questionnaire}.
Note that for question 8, lower scores indicate better results, as the question is phrased so that agreement corresponds to a less desirable response.
For all other questions, higher indicates better results.
An ANOVA test comparing participants' ASA scores indicated significant differences among groups: $F=4.15, p=.018$.
Tukey's Honestly Significant Difference showed that the difference lies between 8B and 70B, favouring the 70B model.
Because of this difference, we zoomed in on the individual questions for both (8B and 70B) groups.
Overall, the 70B model was rated highest in helpfulness, usability, and reliability, and users had the most positive attitude towards it.
The 3B scored best on the subjective engagement and responsiveness scores and shares the highest rank in likeability with the 70B model. 
The 8B model was perceived most as a social entity.
To test for differences across all conditions, a Kruskal-Wallis test was used.
Significant differences were found across all three conditions for questions 4, 5, and 8.
Yet, after Dunn's post hoc test, only the difference in reliance between the 8B and 70B conditions remained, showing the 70B model was perceived as more reliable.
Hence, we find no appreciable differences in perceived usability and satisfaction between model sizes (\textbf{RQ3}).


\paragraph{Additional qualitative analysis}  
Inspection of the log of the exact phrasings that participants used in the chat window revealed that the most prevalent strategy was copy-pasting the question to the RAG-assistant and pressing return.
For some questions, this yields an accurate result; however, other questions needed another strategy to obtain the correct answer.
When this strategy did not work, users diverted to other, more interactive strategies.
Following what \citep{zhou2024developing} have termed `distribution', participants split the question into sub-questions, which were asked separately. 
The most prevalent follow-up was providing more context to the assistant. 
Contextualisation often occurred after participants themselves found and provided relevant excerpts to the assistant.
Task evolution \citep{zhou2024developing} was observed when participants started with a base question and progressively sought the right answer.
A collaboration strategy that occurred less often but stood out was, for example, `backtracking' from the answer options provided in the three multiple-choice questions.

Zooming in on performance per question, we observed that participants scored poorest on Question 5: \textit{`How long does it take to turn a plane around in the air at standard rate?'}
A likely source of difficulty was ambiguity in the interpretation of `turning around'. 
Analysis of the interactions revealed that the AI assistant frequently returned information regarding a 360\degree{} turn, rather than a 180\degree{} turn. 
Conversely, even when participants were presented with information explicitly referring to a 360\degree{} turn, they often failed to recognise that this did not correspond to `turning around', and consequently reported incorrect answers.

\section{Discussion} 
Our results highlight that human-AI collaboration yields significantly higher performance than LLM-only and LLM+RAG systems.
The 3B and 8B models benefit more from human-AI interaction than the 70B model. 
Within the Human-AI setting, task accuracy increased to the degree that the performance gap between the 8B and 70B models was closed. 
While size matters for LLM+RAG settings, human-AI collaboration mitigates this difference, corroborating the synergetic nature of hybrid intelligence \cite{akata2020research}.

The ASA scores reflected greater appreciation among participants for the 70B model than for the 8B model. 
Subjective ratings showed significant differences only in reliability and affective evaluation, in favour of the 70B model.
For all other measures, participants rated the assistants similarly.
A similar pattern was reported by \citep{de2024evaluating}, who found that the smaller Llama model in their setup demonstrated similar or even better general alignment with users than its larger counterpart. 
Individual preferences played a role as well: for some questions, 3B scored higher than 70B and vice versa, visible in \autoref{tab:ai_questionnaire}. 

Although the 8B model outperforms the 3B model on benchmarks involving language understanding and reasoning \cite{suzgun-etal-2023-challenging}, the subjective ratings for the 8B model were lower than those for the 3B and 70B models. 
Specifically, users found the 8B model less reliable than the 70B model.
This might be attributed to its worst performance on instruction-following benchmarks \cite{zhou2023instruction}, or because of the version difference between the two Llama versions. 
Consistent with \citep{fitrianie2022artificial}, we observed correlations between certain questionnaire items (e.g., agent's attitude and agent's likeability).
Such correlations stem from lexical ambiguity or conceptual similarity between the items. 
While this does not invalidate our findings, it suggests that some responses might capture overlapping aspects of participants’ attitudes.
In conclusion, participants indicate differences in some areas, but these do not amount to a decisive preference for one of the three models. 

Overall, the 8B and 70B models yield the highest objective accuracy.
We observe a disconnect between accuracy and perceived usability and satisfaction, as the better-performing models were not rated higher than the 3B model in most areas.
Understanding this disconnect could improve not only evaluation frameworks but also the design of user-facing systems that prioritise trust, interpretability, and perceived usefulness alongside performance.
This suggests a need for deeper exploration of subjective user evaluations.

In relation to previous work, \citep{vaccaro2024humanai} showed that the added value of human-AI collaboration is task-dependent.
While a case for the value of hybrid setups has been made for classification \cite{ma2023should} and deliberation \cite{vandermeer2022hyena}, our work suggests that human-AI collaboration also adds value for QA settings. 
Our hybrid setup can be seen as multi-turn \textit{human} rephrasing, a hybrid variant of the computationally costly strategy of automated rephrasing that is effective \cite{wang-etal-2023-query2doc}.
We suggest that a hybrid setup is a viable and efficient alternative, especially in sensitive, high-stakes, or domain-specific contexts where human involvement is already required.
Especially because with this setup, one could combine the human with a smaller LLM.
Still, several small open-source models could be embedded in a system simulating multi-turn rephrasing. 
This could test whether small models can effectively handle sensitive real-world data, making it an interesting future direction.
As a final point, we observed that the standard deviation for the 3B model is larger than for the 8B and 70B models in the human-AI collaboration setting.
Smaller models are more sensitive to input changes than larger models \cite{huang2024survey, shao2024scaling}.
Lastly, ensuring these results generalise broadly, it has to be tested with models from different families.

As for the limitations of our setup, the fixed retriever returns chunks in similar orders for similar queries. 
Given that LLMs exhibit primacy and recency effects \cite{liu-etal-2024-lost}, chunk position in the context may have influenced performance.
This was partly mitigated by participants generating their own queries, which resulted in chunks presented in different orders. 
The analyses presented in \autoref{sec:quantRes} also account for such kinds of variability by incorporating both participants and questions as random effects in the mixed-effect models.
Additionally, participants did not receive feedback on the correctness of their answers: while automated evaluation would have been feasible for multiple-choice questions, open-ended responses required manual assessment. 
This lack of feedback may have shaped perceptions of the models’ helpfulness.
On the other hand, in real workplace settings, the correct answer would often not be available either. Additionally, participants could review and reflect on their own answers and were incentivised to do so through potential bonus payments. 
As recruitment was via Prolific, where financial compensation is a primary motivator, some participants may have prioritised speed over accuracy despite these incentives, potentially reducing answer quality and increasing susceptibility to hallucinations. 
In our current setup, we deliberately chose to use only one document that contains a mixture of technical and procedural information. 
In future follow-ups, multiple documents with varying content types could be used. 
Perceived slowness in response times may have resulted from participants' internet connections or latency on the Hugging Face API servers, potentially impacting their experience and perceptions of the AI tools.


\section{Conclusion}
This paper examined human-AI interaction in a realistic information-seeking task mirroring real-world information processing contexts.
It focused on evaluating multiple perspectives: accuracy, usability, and satisfaction across three different conditions, using a combination of performance and participant evaluations. 
The results underscore the importance of human–AI collaboration, which consistently outperformed LLM-only and LLM+RAG in our experiment. 
Specifically, smaller models benefitted more from human collaboration to the degree that this closed the performance gap between the 8B and 70B models.
Importantly, participants did not perceive larger models as more capable than smaller models, revealing a disconnect between benchmark performance and human user experience.
This asks for looking beyond pure benchmark performance, assessing applications in realistic, multi-turn scenarios that take user perspectives into account.

\bibliographystyle{vancouver}
\bibliography{hhai}

\end{document}